# Beampattern Design in Non-Uniform MIMO Communication

Amirsadegh Roshanzamir
*Department of Information Systems and Management*
*University of South Florida*
roshanzamir@usf.edu

**ABSTRACT**

In recent years and with introduction of 5G cellular network and communication, researchers have shown great interest in Multiple Input Multiple Output (MIMO) communication, an advanced technology. Many studies have examined the problem of designing the beampattern for MIMO communication using uniform arrays and the covariance-based method to concentrate the transmitted power to the users. However, this paper aims to tackle this issue in the context of non-uniform arrays. Previous authors have primarily focused on designing the transmitted beampattern based on the cross-correlation matrix of transmitted signal elements. In contrast, this paper suggests optimizing the positions of transmitted antennas along with the cross-correlation matrix. This approach is expected to produce better results.

## I. INTRODUCTION

MIMO communication, a burgeoning field in the realm of communication, has recently captivated researchers worldwide. Unlike conventional phased array communication and antenna design, MIMO communication offers the freedom to select transmitted probing signals, enabling the maximization of power around specific user locations of interest [1, 2]. These signals can either be correlated or non-correlated, providing additional degrees of freedom for communication system design.

In the context of MIMO communication, particularly in 5G networks, colocated antennas are utilized on the transmitter side, with antennas situated in close proximity to one another. This type of MIMO communication leverages the concept of MIMO to enhance spatial resolution. Numerous studies have showcased the benefits of this approach, including superior interference rejection capability [4, 5], enhanced parameter identifiability [6], and increased flexibility for transmit beampattern design [7, 8]. Valuable resources on MIMO communication with colocated antennas can be found in [3].

Typically, covariance matrix based design methods serve as the primary approach for MIMO communication waveform design [7] – [11]. These methods focus on the cross correlation matrix of transmitted signals rather than the entire waveform, thereby impacting solely the spatial domain. References [7, 9] emphasize designing the cross correlation matrix of transmitted signals to facilitate power transmission within a desired range of angles. In [8], the cross correlation matrix of transmitted signals is tailored to enable control over spatial power. Additionally, [8] minimizes the cross correlation between transmitted signals at various user locations, thereby enhancing spatial resolution in the communication receiver. The authors in [10] optimize waveform covariance based on the Cramer – Rao bound matrix, while [11] designs corresponding signal waveforms to achieve low peak to average power ratio (PAR) and improved user resolution, given the optimized covariance matrix.

However, all the aforementioned methods solely explore transmit beampattern design using a uniform linear array (ULA) with half-wavelength inter-element spacing. In reality, the selection of array position can offer additional degrees of freedom for transmit beampattern matching design. Consequently, a joint optimization of the cross correlation matrix of transmitted signals and antenna location can yield superior outcomes compared to existing methods employing a ULA.

This study comprises six sections. Section 1 provides a concise introduction to MIMO communications. Section 2 delves into the review of covariance-based MIMO communication beamforming. Section 3 discusses the utilization of a non-uniform beampattern matching design model. Section 4 focuses on solving the optimization problem. Numerical examples are presented in section 5. Finally, section 6 concludes the paper, and references are provided at the end.

## II. COVARIANCE BASED BEAMPATTERN DESIGN

Let's consider a scenario where we possess a set of N transmitting antennas positioned at known coordinates $\boldsymbol{x_i} = (x_{1,i}, x_{2,i}, x_{3i}) = (x, y, z)$ in a spherical coordinate system along the z-axis, as depicted in Fig. 1. It is important to note that for the purpose of this study, we are focusing on non-uniform arrays, meaning that the distances between these N transmitting antennas are completely arbitrary. Throughout this research and in all instances of examples and formulas mentioned in this paper, we assume that these transmitting antennas are aligned along the z-axis.

We make the assumption that each transmitting antenna is activated by a specific signal at the carrier frequency $f_c$ or with a wavelength of $\lambda$, possessing a complex envelope denoted as $s_i(t)$, where i ranges from 1 to N. At any given point in space, located at a distance r and with a direction represented by $k(\theta, \phi)$ relative to the transmitting antenna, the radiated signal in the far field is characterized by a complex envelope, which can be expressed as follows.

$$y_i(t, r, \theta, \phi) = \frac{1}{\sqrt{4\pi r}} s_i\left(t - \frac{r}{c}\right) e^{j\left(\frac{2\pi}{\lambda}\right) x_i^T k(\theta, \phi)} \quad (1)$$

Where, in this equation, $k$ is a unit vector in the $(\theta, \phi)$ direction.

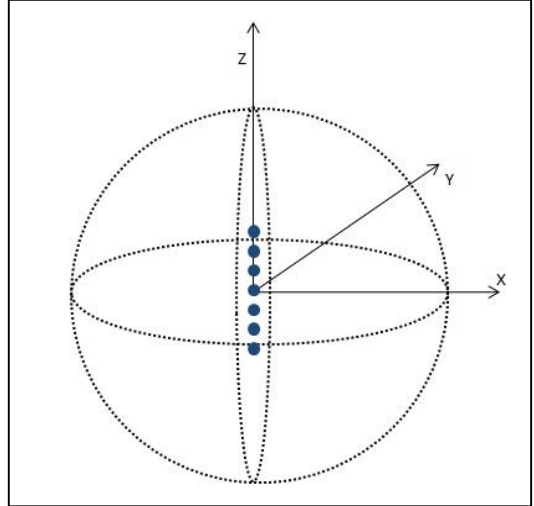

**Fig. 1. Transmitter antennas and spherical coordinate system**

At the far field, these signals add linearly and the radiated powers $P_i$ add linearly as well. At this point assume that the i-th element location is on the z-axis at coordinate $z_i$. The signal at position $(r, \theta, \phi)$ resulting from all of the transmitted signals at far field will be:

$$y(t, r, \theta, \phi) = \sum_{i=1}^{N} y_i(t, r, \theta, \phi) = \frac{1}{\sqrt{4\pi r}} \sum_{i=1}^{N} s_i\left(t - \frac{r}{c}\right) e^{j\left(\frac{2\pi z_i}{\lambda}\right) \sin(\theta)} \quad (2)$$

The power density of the entire signals then given by

$$P_y(r, \theta, \phi) = \mathbb{E}\{|y(t, r, \theta, \phi)|^2\} = \frac{1}{4\pi r^2} \sum_{k=1}^{N} \sum_{l=1}^{N} <s_k(t) s_l^*(t)> \, e^{j\left(\frac{2\pi(z_k - z_l)}{\lambda}\right) \sin(\theta)} \quad (3)$$

Where $\mathbb{E}\{\}$ is statistical expected value.
It is known that the complex signal cross-correlation is defined by
$$R_{kl} = <s_k(t) s_l^*(t)> \quad (4)$$

With defining the direction vector as below
$$\boldsymbol{a}(\theta) = \left[e^{j\left(\frac{2\pi z_1}{\lambda}\right)\sin(\theta)}, \ldots, e^{j\left(\frac{2\pi z_N}{\lambda}\right)\sin(\theta)}\right]^T \quad (5)$$
The normalized power density $P(\theta,\phi)$ of signals, in (W/ster), would be:
$$P(\theta,\phi) = \frac{1}{4\pi}\sum_{k=1}^{N}\sum_{l=1}^{N} R_{kl} e^{\frac{j2\pi}{\lambda}(z_k - z_l)\sin(\theta)} \quad (6)$$
Recognizing that (6) is quadratic form in the Hermitian Matrix R which is the cross-correlation matrix of signals, this can be written compactly as
$$P(\theta,\phi) = \frac{1}{4\pi}\mathbf{a}^*(\theta)\mathbf{R}\mathbf{a}(\theta) = \frac{1}{4\pi}\Re\{\mathbf{R}\odot(\mathbf{a}(\theta)\mathbf{a}^*(\theta))\} \quad (7)$$
It should be mentioned that in this equation $\Re\{\}$ is a real part of a complex vector and $\odot$ denotes the inner product of matrices. This normalized power density $P(\theta,\phi)$ in (7) is exactly the beampattern which is desirable to find.

In the following some examples of beampatterns produce from such a cross-correlation matrix has been shown. Fig. 2 shows the beampattern produced by signal cross-correlation matrix of (8), (9) and (10) respectively. It is noticeable that these figures are beam-patterns of 10-element uniform linear array (ULA) with half-wavelength spacing (in general case we want to consider Non Uniform Linear Arrays(NULA)).

$$\begin{bmatrix} 1 & \cdots & 1 \\ \vdots & \ddots & \vdots \\ 1 & \cdots & 1 \end{bmatrix} \quad (8)$$

$$\begin{bmatrix} 0.8^0 & \cdots & 0.8^9 \\ \vdots & \ddots & \vdots \\ 0.8^9 & \cdots & 0.8^0 \end{bmatrix} \quad (9)$$

$$\begin{bmatrix} 1 & \cdots & 0 \\ \vdots & \ddots & \vdots \\ 0 & \cdots & 1 \end{bmatrix} \quad (10)$$

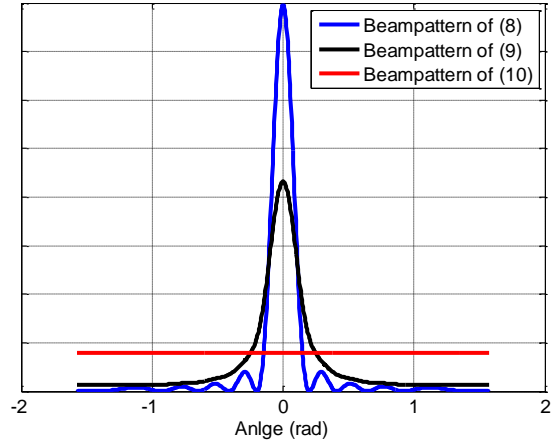

**Fig. 2. Beampattern respect to (7). The blue one is corresponds to cross-correlation matrix of (8), the black one is corresponds to cross-correlation matrix of (9) and the red one is corresponds to cross-correlation matrix of (10)**

In general case the elements of the signal cross-correlation matrix are complex values except the diagonal elements that are real. This general case is related to MIMO communications but in the case of phased array communication, all the transmitter signals are correlated with each other and then absolute value of all the elements in $R$, are equal to 1(blue one at Fig. 2).

### III. NON UNIFORM MIMO COMMUNICATION BEAMPATTERN MATCHING DESIGN

In this section the problem of beampattern matching design for a non-uniform array will be considered. Mathematically, the cost function of the problem for concentrating power to K users, can be formulated as follow:

$$J(\mathbf{R}) = \frac{1}{K}\sum_{k=1}^{K}\omega_k\left(\Re\{\mathbf{R}\odot(\mathbf{a}(\theta)\mathbf{a}^*(\theta))\} - \mathbf{P}_d(\theta_k)\right)^2 \quad (11)$$

Where, $\omega_k \geq 0$ is the weight for the $\theta_k$. $\mathbf{P}_d(\theta)$ denotes the desired beampattern.

Since $\mathbf{R}$ is a covariance matrix, it must be a positive semidefinite matrix. Moreover, if all the antennas are required to transmit the uniform power, all diagonal elements of $R$ must be the same:

$$\mathbf{R} \geq 0$$
$$\mathbf{R}_{nn} = c, for\ n = 1, \dots, N \quad (12)$$

Where $c$ is a given constant denotes total power.

Now, for considering non-uniform array and without loss of generality and work in discrete domain, we assume that antenna transmitters can be placed in N positions out of M available positions which obviously $M \geq N$. Fig. 3 shows the non-uniform array presentation as a grid line:

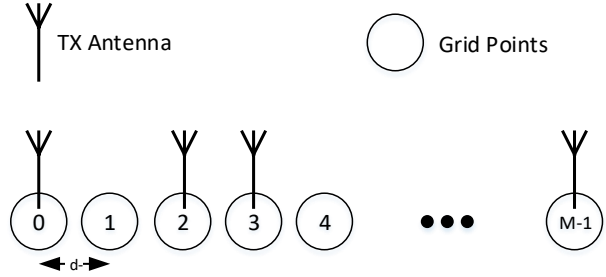

**Fig. 3. Geometry and placement of antennas**

Now, mathematically we can introduce an antenna position vector to represent the antenna configuration scheme:

$$\boldsymbol{g} = [g_1, g_2, \dots, g_M]^T, \qquad g_i \in \{0,1\} \quad (13)$$

Where $g_i$ indicates whether or not the $i$-th grid point is selected then $g_i = 1$, otherwise $g_i = 0$.

And therefore, the new steering vector in (5) can be written as follow:

$$\mathbf{b}(\theta) = (\boldsymbol{g}\odot\mathbf{a}(\theta))^* \quad (14)$$

And the cost function in (11) can be rewrite as follow:

$$J(\boldsymbol{g}, \mathbf{R}) = $$
$$\frac{1}{K}\sum_{k=1}^{K}\omega_k\left(\Re\{\mathbf{R}\odot(\mathbf{b}(\theta)\mathbf{b}^*(\theta))\} - \mathbf{P}_d(\theta_k)\right)^2 =$$
$$\frac{1}{K}\sum_{k=1}^{K}\omega_k\left(\Re\{\mathbf{R}\odot((\boldsymbol{g}\odot\mathbf{a}(\theta))(\boldsymbol{g}\odot\mathbf{a}(\theta))^*)\} - \mathbf{P}_d(\theta_k)\right)^2 =$$
$$\frac{1}{K}\sum_{k=1}^{K}\omega_k\left(\boldsymbol{g}^T\Re\{\mathbf{R}\odot(\mathbf{a}(\theta)\mathbf{a}^*(\theta))^*\}\boldsymbol{g} - \mathbf{P}_d(\theta_k)\right)^2 \quad (15)$$

Therefore, the optimization problem can be formulated as follows:

$$\min_{\boldsymbol{g},\mathbf{R}} J(\boldsymbol{g}, \mathbf{R})$$
$$s.t.\ g_m \in \{0,1\}, \qquad m = 1, \dots, M$$
$$\mathbf{1}^T\boldsymbol{g} = N$$
$$\mathbf{R}_{mm} = c, \qquad m = 1, \dots, M$$
$$\mathbf{R} \geq 0 \quad (16)$$

The formulated problem (16) can be classified as a mixed Boolean-nonconvex problem, since the elements of $\boldsymbol{g}$ must be Boolean so there is no closed-form solution to problem (16).

## IV. SOLUTION TO THE OPTIMIZATION PROBLEM

In this section, we adopt an iterative method to obtain an optimized solution to problem (16).

**IV.I Optimization of $R$**

For a fixed $p$, we optimize $R$ by:
$$\min_{\alpha, \mathbf{R}} J(g, \mathbf{R})$$
$$\mathbf{R}_{mm} = c, \quad m = 1, \dots, M$$
$$\mathbf{R} \geq 0 \tag{17}$$

It is noticed that problem (17) can be formulated as a semidefinite quadratic programming (SQP) [12], which is a convex problem and can be effectively solved via the convex optimization toolbox CVX [13].

**IV.II Optimization of $g$**

For a fixed $R$, we can optimize $p$ by solving the following Boolean optimization problem:
$$\min_{\alpha, \mathbf{R}} J(g, \mathbf{R})$$
$$s.t. \ g_m \in \{0, 1\}, \quad m = 1, \dots, M$$
$$\mathbf{1}^T g = N \tag{18}$$

As in real world of antennas, the antenna spacing is related to wavelength, therefore, here, we replace the nonconvex Boolean constraints $g_m \in \{0,1\}$ with convex constraints $g_m \in [0,1]$ to obtain the relaxed solution of the antenna optimization problem [14], [15].

To simplify notations, $J(g, \mathbf{R})$ in (15) can be expressed as a concise form:
$$J(g, \mathbf{R}) = \frac{1}{K} \sum_{k=1}^{K} \omega_k (\mathbf{x}^T A_k \mathbf{x})^2 \tag{19}$$

Where
$$\mathbf{x} = \begin{bmatrix} \sqrt{\alpha} \\ g \end{bmatrix} \tag{20}$$

And
$$A_k = \begin{bmatrix} -P_d(\theta_k) & 0 \\ 0 & \Re\{\mathbf{R} \odot (\mathbf{a}(\theta)\mathbf{a}^*(\theta))^*\} \end{bmatrix} \tag{21}$$

Consequently, the relaxed problem can be expressed as follow:
$$\min_{\mathbf{x}} \frac{1}{K} \sum_{k=1}^{K} \omega_k (\mathbf{x}^T A_k \mathbf{x})^2$$
$$s.t. \ 0 \leq x_i \leq 1, \quad i = 2, \dots, M+1$$
$$\mathbf{1}^T \mathbf{x}(2:M+1) = N \tag{22}$$

It is apparent that the goal function in (22) is a non-linear fourth-degree polynomial, which presents a challenging NP-hard dilemma [16]. In this scenario, we employ the ADMM technique [17] to discover a resolution for problem (22). We break down the fourth-degree polynomial issue into a two-variable quadratic problem by introducing a primal variable $\mathbf{y}$. Consequently, problem (22) can be transformed into the subsequent problem:
$$\min_{\mathbf{x}} \frac{1}{K} \sum_{k=1}^{K} \omega_k (\mathbf{x}^T A_k \mathbf{x})^2$$
$$s.t. \ \mathbf{x} = \mathbf{y}$$
$$\mathbf{x} \in \chi_x, \quad \mathbf{y} \in \chi_y \tag{23}$$

Where the convex sets $\chi_x$ and $\chi_y$ are respectively:
$$\chi_x = \left\{ \mathbf{x} \middle| \begin{array}{l} 0 \leq x_i \leq 1, \ i = 2, \dots, M+1 \\ \mathbf{1}^T \mathbf{x}(2:M+1) = N \end{array} \right\} \tag{24}$$

And
$$\chi_y = \left\{ \mathbf{y} \middle| \begin{array}{l} 0 \leq y_i \leq 1, \ i = 2, \dots, M+1 \\ \mathbf{1}^T \mathbf{y}(2:M+1) = N \end{array} \right\} \tag{25}$$

To simplify the analysis, we define:
$$F(\mathbf{x}, \mathbf{y}) = \frac{1}{K} \sum_{k=1}^{K} \omega_k (\mathbf{x}^T A_k \mathbf{x})^2 \tag{26}$$

We form the augmented Lagrangian (using the scaled form) of problem (26) as follows:
$$\mathcal{L}(\mathbf{x}, \mathbf{y}, \mathbf{u}) = F(\mathbf{x}, \mathbf{y}) + \frac{\rho}{2} \|\mathbf{x} - \mathbf{y} + \mathbf{u}\|^2 \tag{27}$$

Where $\rho \geq 0$ is a penalty parameter, and $\mathbf{u}$ is a scaled dual variable.

Therefore, at $(m+1)$-th iteration, he ADMM consists of the iterations [17]

$$\mathbf{x}^{m+1} := \arg\min_{\mathbf{x} \in \chi_x} \mathcal{L}(\mathbf{x}, \mathbf{y}^m, \mathbf{u}^m) \quad (28)$$

$$\mathbf{y}^{m+1} := \arg\min_{\mathbf{y} \in \chi_y} \mathcal{L}(\mathbf{x}^{m+1}, \mathbf{y}, \mathbf{u}^m) \quad (29)$$

$$\mathbf{u}^{m+1} := \mathbf{u}^m + \mathbf{x}^{m+1} - \mathbf{y}^{m+1} \quad (30)$$

It is quite easy to confirm that issues (28) and (29) are Quadratic Programmings (QPs), which are problems with convexity. In this case, we can acquire their resolutions by addressing the linked Karush-Kuhn-Tucher (KKT) prerequisites [16].

Specifically, the solution $x^*$ for problem (22) might be fractional. Hence, we can choose the N nearest components of $x^*$, resulting in a suboptimal solution $\hat{x}$ for problem (18) [14], [18].

## V. NUMERICAL SIMULATIONS

In this section, we assess the performance of the proposed method through numerical simulations. Unless specified otherwise, all simulations assume a Uniform Linear Array (ULA) with 65 grid points and one-eighth wavelength inter-grid interval. The number of transmit antennas is set to 15. The angle range is (-90°, 90°) with a spacing of 1°, and the weight for each angle is equal for all the users. For the penalty parameter $\rho$ used in the optimization of $\mathbf{g}$, we choose $\rho = 30$. The initial antenna position $g_0$ is selected to be a traditional uniform array.

We consider a desired beampattern with two mainlobes at $\theta_1 = 40°$ and $\theta_2 = -20°$, with a beamwidth of $\Delta\theta_1 = 25°$ and $\Delta\theta_2 = 10°$. Fig. 4 illustrates the values of the objective function in problem (8), which represents the Mean Squared Error (MSE) between the desired beampattern and the designed beampattern, plotted against the iteration number. The results in Fig. 4 demonstrate that the MSE values decrease as the iteration number increases, indicating that the proposed algorithm converges rapidly (i.e., after 10 iterations). As expected, the Optimized Array achieves a lower MSE level compared to the uniform array, with a level gap of approximately 6. This implies that the non-uniform array can achieve a better beampattern matching performance.

Next, Fig. 5 compares the resulting beampattern of the non-uniform array with that of the uniform array. The results show that the optimized array exhibits comparable and superior sidelobe performance compared to the conventional uniform array. This suggests that jointly optimization of the covariance matrix R and antenna position $\mathbf{g}$ can yield better results than optimizing only the covariance matrix R with the uniform array. Additionally, Fig. 6 displays the antenna position of the non-uniform obtained from the proposed algorithm. Interestingly, the effective aperture of the non-uniform is found to be around 30, indicating that increasing the number of candidate grid points M beyond 30 does not provide additional degrees of freedom to match the desired beampattern when M > 30 and N = 15 in this particular example.

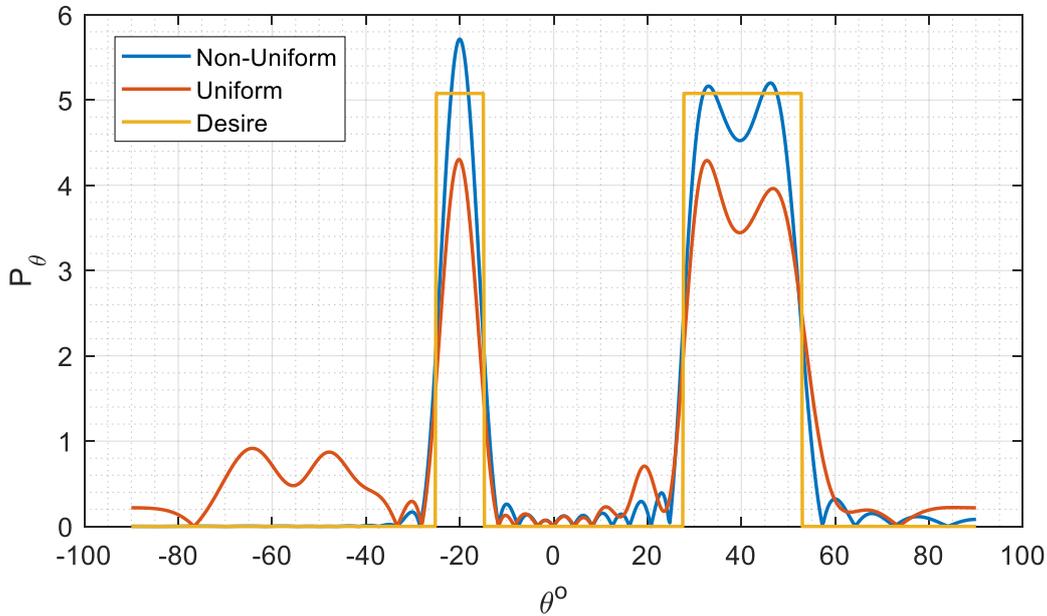

**Fig. 4. Resulted non-uniform beampattern compare to uniform array**

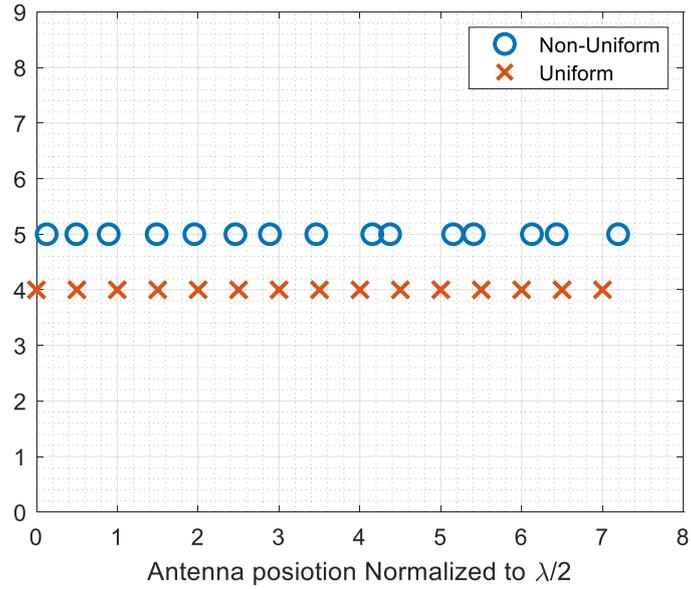

**Fig. 5. Non-uniform antenna posiotion compare to uniform for M = 65 and N = 15**

Furthermore, Fig. 7 presents the values of the objective function in problem (16) plotted against the number of grid points M. The results reveal that as M increases up to 30, the degree of freedom for beampattern design also increases.

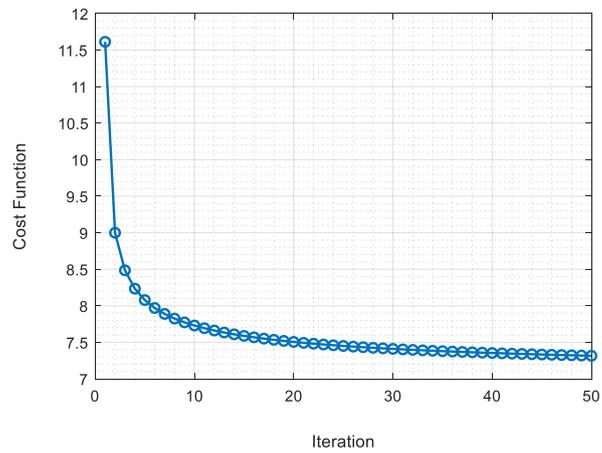

**Fig. 4. The values of cost function in problem (16) versus the iteration number. M=65, N=15**

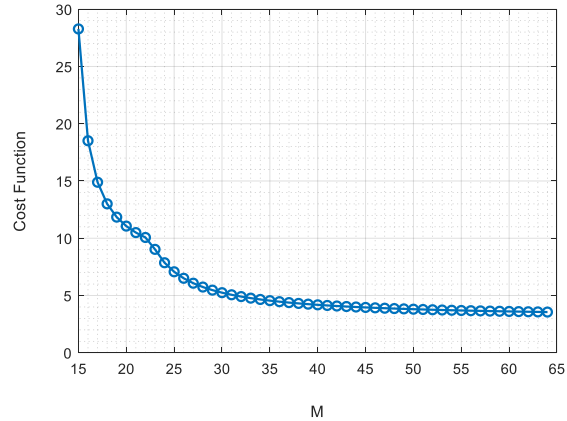

Fig. 6. The values of cost function in problem (16) versus the number of the grid points M with N=15

## VI. CONCLUSION

This paper focuses on addressing the issue of optimizing the covariance matrix **R** and the position of the antenna **g** for MIMO communication transmit beampattern matching design. To tackle this problem, which involves a nonconvex objective function and a Boolean-nonconvex constraint, we propose an iterative approach to jointly optimize **R** and **g**. In each iteration, **R** is determined using the SQP method, and the relaxed solution of **g** is obtained through the ADMM algorithm.

The results from numerical simulations demonstrate that the optimization of **R** and **g** together leads to improved mainlobe matching performance and reduced sidelobe levels compared to solely optimizing **R** with a uniform array.